\documentclass[12pt]{article}

\begin{document}

\begin{center}
{\bf Squeezing in Multivariate Spin Systems}
\end{center}

\begin{center}
{\bf Swarnamala Sirsi}
\end{center}
\begin{center}
Department of Physics,\\ Yuvaraja's College, University of Mysore\\
Mysore 570005, India email: ssirsi@uomphysics.net
\end{center}

\begin{abstract}
In contrast to the canonically conjugate variates $q$,$p$ representing the position and momentum of a particle in the phase space distributions, the three Cartesian components, $J_{x}$,$J_{y}$, $J_{z}$ of a spin-$j$ system constitute the mutually non-commuting variates in the quasi-probabilistic spin distributions.  It can be shown that a univariate spin distribution is never squeezed and one needs to look into either bivariate or trivariate distributions for signatures of squeezing.  Several such distributions result if one considers different characteristic functions or moments based on various correspondence rules. As an example, discrete probability distribution for an arbitrary spin-$1$ assembly is constructed using Wigner-Weyl and Margenau-Hill correspondence rules. It is also shown that a trivariate spin-$1$ assembly resulting from the exposure of nucleus with non-zero quadrupole moment to combined electric quadrupole field and dipole magnetic field exhibits squeezing in cerain cases. 
\end{abstract}

\begin{center}
{\bf keywords:}Multivariate spin distributions, spin squeezing, entanglement
\end{center}

\section{Introduction}
A spin-$j$ system can be represented by an $n \times n$ density matrix where $n=2j+1.$ Any spin-$j$  density matrix can be parametrized\cite{fano} in terms of $n^{2}-1$ independent, spherical tensor parameters $t^{k}_{q}$
as
\begin{equation}
\rho ={1\over 2j+1}\sum _{k=0}^{2j}\sum_{q=-k}^{+k} t^{k}_{q}{\tau^{k}_{q}}^{\dagger}(\vec J)
\end{equation}
where 
\begin{equation}
t^{k}_{q}=Tr(\rho \tau^{k}_{q}(\vec J)).
\end{equation}
%\end{document}
Spherical tensor operators $\tau^{k}_{q}(\vec J)$ can be obtained by Weyl's construction\cite{rose} using spin operators $J_{x},J_{y},J_{z}$ as
\begin{equation}
\tau_{q}^{k}(\vec J) = N^{-1}_{kj}(\vec J\cdot \vec \nabla)^{k} r^{k} Y_{q}^{k}(\theta, \phi)
\end{equation}
where $ N^{-1}_{kj}$ are the normalization factors given by
\begin{equation}
N^{-1}_{kj} = \frac{2^{k}}{k!}\sqrt{\frac{4\pi(2j-k)!(2j+1)}{(2j+k+1)!}}
\end{equation}
which are in agreement with the earlier usage of Ramachandran and Umergee\cite{gr1}  and with the Madison convention\cite{sach}.
%\end{document}
Here $Y^{k}_{q}(\theta , \phi )$ are the well known spherical harmonics.  The spherical tensor operators  satisfy the orthogonality relations,
\begin{equation}
Tr(\tau_{q}^{k}({\vec J})\tau_{q'}^{k'^{\dagger}}({\vec J}))=n\delta _{kk^{'}qq^{'}}
\end{equation}
where
\begin{equation}
\tau_{q}^{k^{\dagger}}(\vec J)=(-1)^{q} \tau_{-q}^{k}(\vec J).
\end{equation}
Hermiticity of the density matrix together with Eq.(6) demands that
\begin{equation}
t_{q}^{k^{*}}=(-1)^{q} t^{k}_{-q}.
\end{equation}
The set of $n^{2}-1$ spherical tensor operators $\tau^{k}_{q}(\vec J)$ together with the identity matrix, form an orthonormal basis of the linear vector space of complex $n\times n$  matrices that act on the $n$ dimensional spin space.
 Thus $\rho$ may be completely determined by finding the expectation values of these $n^{2}-1$ operators in the state $\rho.$
 
Consider the example of a spin-$1$ density matrix. It can be completely characterized by the spherical tensor parameters 
$t^{1}_{0},t^{1}_{\pm 1},t^{2}_{0},t^{2}_{\pm 1},t^{2}_{\pm 2}$
and the corresponding spherical tensor operators, 
$\tau ^{1}_{q}$ and $\tau^{2}_{q}$ can be written as
\begin{equation}
\tau^{1}_{q}(\vec J)=N_{1j}^{-1}(\vec J\cdot \vec \nabla)rY^{1}_{q}(\theta,\phi)
\end{equation}
and
\begin{equation}
\tau^{2}_{q}(\vec J)=N_{2j}^{-1}(\vec J\cdot \vec \nabla)^{2}r^{2}Y^{2}_{q}(\theta,\phi) .
\end{equation}
Explicitly,
\begin{equation}
\tau_{0}^{1}=N_{1j}^{-1} (J_{x}\frac{\partial}{\partial x}+J_{y}\frac{\partial}{\partial y}+J_{z}\frac{\partial}{\partial z})(\sqrt{\frac{3}{4\pi}}z)
\end{equation}
\begin{equation}
\tau_{\pm 1}^{1}=\mp N_{1j}^{-1} (J_{x}\frac{\partial}{\partial x}+J_{y}\frac{\partial}{\partial y}+J_{z}\frac{\partial}{\partial z})\sqrt{\frac{3}{8\pi}}(x\pm iy)
\end{equation}
\begin{equation}
\tau_{0}^{2}=N_{2j}^{-1} (J_{x}\frac{\partial}{\partial x}+J_{y}\frac{\partial}{\partial y}+J_{z}\frac{\partial}{\partial z})^{2}\sqrt{\frac{5}{16\pi}}(3z^{2}-1)
\end{equation}
\begin{equation}
\tau_{\pm 1}^{2}=\mp N_{2j}^{-1} (J_{x}\frac{\partial}{\partial x}+J_{y}\frac{\partial}{\partial y}+J_{z}\frac{\partial}{\partial z})^{2}\sqrt{\frac{15}{8\pi}}[z(x\pm iy)]
\end{equation}
\begin{equation}
\tau_{\pm 2}^{2}=N_{2j}^{-1} (J_{x}\frac{\partial}{\partial x}+J_{y}\frac{\partial}{\partial y}+J_{z}\frac{\partial}{\partial z})^{2}\sqrt{\frac{15}{32\pi}}(x^{2}-y^{2}\pm 2ixy)
\end{equation}
and
\begin{eqnarray}
t^{1}_{0}=&N_{1j}^{-1} \sqrt{\frac{3}{4\pi}}Tr{ \rho J_{Z}}\\
t^{1}_{\pm 1}=& \mp N_{1j}^{-1}\sqrt{\frac{3}{8\pi}}Tr(\rho( J_{X}\pm i J_{Y}))
\end{eqnarray}
\begin{eqnarray}
t^{2}_{0}=& N_{2j}^{-1} \sqrt{\frac{5}{16 \pi}}Tr(\rho ( 3J_{Z}^{2}-j(j+1)))\\\
t^{2}_{\pm 1}=&\mp N_{2j}^{-1} \sqrt{\frac{15}{8\pi}}Tr(\rho( (J_{Z}J_{X}+J_{X}J_{Z})\pm i (J_{Z}J_{Y}+J_{Y}J_{Z})))\\
t^{2}_{\pm 2}=& N_{2j}^{-1} \sqrt{\frac{15}{32\pi}}Tr(\rho( (J_{X}^{2}-J_{Y}^{2})\pm 2i (J_{X}J_{Y}+J_{Y}J_{X}))).
\end{eqnarray}
Here one could easily identify $t^{1}_{q},t^{2}_{q}$ as linear combinations of $1^{st}$ and $2^{nd}$ order Cartesian moments of the operators  $J_{X},J_{Y},J_{Z}.$ 
Further, computing all the spherical tensor parameters characterising a spin-$j$ density matrix, one could see that all the  $t^{k}_{q}$'s can be constructed by taking appropriate linear combinations of the $k^{th}$ order cartesian mixed moments $\mu^{abc},(a+b+c=k)$ of the operators $J_{X},J_{Y},J_{Z}$  provided the mixed moments are defined by the relation
\begin{equation}
\mu^{abc}=\frac{a!b!c!}{(a+b+c)!}Tr [\rho{\hbox{S}_{ww}}(J_{X}^{a}J_{Y}^{b}J_{Z}^{c})]
\end{equation}
where $S_{ww}()$ denotes the the complete symmetrizer, also known as Wigner-Weyl symmetrizer, with $\frac{(a+b+c)!}{a!b!c!}$ terms. In classical probability theory mixed moments are defined through characteristic function $ \Phi(\vec I)$ as 
\begin{equation}
\mu^{abc}=(-i)^{a+b+c}\frac{\partial^{a}}{\partial I_{x}^{a}}\frac{\partial ^{b}}{\partial I_{y}^{b}}\frac{\partial^{c}}{\partial I_{z}^{c}} \Phi (\vec I) \vert _{I_{x}=I_{y}=I_{z}=0}.
\end{equation}
Therefore one can identify the characteristic  function   
leading to the multivariate moments in Eq.(20) to be
\begin{equation}
\Phi_{ww}(\vec I)= Tr(\rho e^{i\vec I\cdot \vec J}).
\end{equation}
Note that this function can be obtained by the  Wigner-Weyl  prescription,
\begin{equation}
e^{i\vec I \cdot \vec X} \rightarrow e^{i \vec I\cdot \vec J}
\end{equation} 
or eqivalently, 
\begin{equation}
X^{a}Y^{b}Z^{c}=\frac{a!b!c!}{(a+b+c)!}\hbox{S}_{ww}(J_{X}^{a}J_{Y}^{b}J_{Z}^{c}).
\end{equation}

Here $X,Y,Z$ are the classical variates corresponding to the quantum mechanical operators $J_{X},J_{Y},J_{Z}$ respectively.
Taking the fourier transform of the characteristic function, $\Phi_{ww}(\vec I)$ one can obtain the { Wigner-Weyl} probability density function\cite{gr2} as
\begin{eqnarray}
 f_{ww}(\vec X) =&\frac{1}{(2\pi)^{3}} \int_{-\infty}^{\infty}\int_{-\infty}^{\infty}\int_{-\infty}^{\infty}d^{3} I \quad \Phi_{ww}(\vec I) e^{-i \vec I \cdot \vec X}\\
 =& \frac{1}{(2j+1)}\sum_{k=0}^{2j}\sum_{q=-k}^{k} (i)^{k} t_{q}^{k*} \nabla^{k}_{q}(\vec X){\cal I}^{k}(X)
 \end{eqnarray}
 where 
 \begin{equation}
 {\cal I}^{k}(X)=\frac{(-1)^{k+1}}{2\pi^{3/2}}\sum_{m=-j}^{j}C(jkj;m0m) \frac{1}{X}\frac{d}{dX} \int_{-\infty}^{\infty} dI \quad I^{-k} e^{iI(X-m)}
 \end{equation}
 and $\nabla_{q}^{k}(\vec X)$ are the spherical components of the gradient operator and $C(jkj;m0m)$ are the well known Clebsch-Gordon coeffecients.
 One could easily verify that 
 \begin{equation}
 {\cal I}_{o}(X) = \frac{-1}{\sqrt{\pi}}\frac{1}{X}\frac{d}{dX}\sum_{m} \delta(X-m)
 \end{equation}
 and
 \begin{equation}
 {\cal I}_{k}(X)=\frac{(-1)^{k+1}(i)^{k}}{(k-1)!\sqrt{\pi}}\sum_{m} C(jkj;m0m) \frac{1}{X}\frac{d}{dX} [(X-m)^{k-1}\epsilon(X-m)]
 \end{equation}
 for $k\ge 1$.  Here $ \delta(X-m)$ denotes the Dirac $\delta-$ function and $\epsilon(X-m)$ denotes the step function
 \begin{equation}
 \epsilon(a)=\frac{1}{2\pi i}\int_{-\infty}^{\infty} dt \frac{e^{iat}}{t}.
 \end{equation}
 Here $\epsilon(a)=1$ if $a> 0$ and $\epsilon(a)=0$ if $a< 0$.\\
 
Thus the problem of finding a quantum mechanical density function reduces to finding the quantum mechanical equivalent of the classical characteristic function or the classical moments. Owing to the noncommutative nature of the spin components $J_{X},J_{Y}$ and $J_{Z},$ one can realize several quantum mechanical definitions of the characteristic function $\Phi (\vec I)$ or mixed moments $\mu ^{abc}$ based on different operator orderings. Once the moments are obtained, instead of a continuous probability distribution function, one can obtain a discrete probability mass function. The procedure for obtaining a discrete probability distribution for a spin-$j$ system is outlined in Sec.2 . Rules associating classical quantities to quantum mechancial operators are called correspondence rules.
 There are many such rules\cite{cohen}. Of these, the Margenau-Hill and the Wigner-Weyl procedures are well studied in the discussion of quantum phase distributions and quantum spin distributions \cite{hillery},\cite{gr2}. As an example we would like to compute the probability mass function for
any general spin-1 system using both Margenau-Hill and Wigner-Weyl correspondence rules in Sec.3.In Sec.4 we discuss the squeezing of univariate and multivariate spin distributions.

\section{Discrete probability distributions of spin-j system}

 Margenau-Hill correspondence rule is given by
 \begin{eqnarray}
 e^{i {\vec I}\cdot {\vec X}} \rightarrow & \frac{1}{3!}[e^{iI_{x}J_{x}}e^{iI_{y}J_{y}}e^{iI_{z}J_{z}}+e^{iI_{x}J_{x}}e^{iI_{y}J_{y}}e^{iI_{z}J_{z}}+e^{iI_{x}J_{x}}e^{iI_{y}J_{y}}e^{iI_{z}J_{z}}\\ \nonumber & +e^{iI_{x}J_{x}}e^{iI_{y}J_{y}}e^{iI_{z}J_{z}}+e^{iI_{x}J_{x}}e^{iI_{y}J_{y}}e^{iI_{z}J_{z}}+e^{iI_{x}J_{x}}e^{iI_{y}J_{y}}e^{iI_{z}J_{z}}]\
 \end{eqnarray}
 or 
\begin{eqnarray}
 X^{a}Y^{b}Z^{c}\rightarrow & \frac{1}{3!}[J_{X}^{a}J_{Y}^{b}J_{Z}^{c}+J_{X}^{a}J_{Z}^{c}J_{Y}^{b}+J_{Y}^{b}J_{X}^{a}J_{Z}^{c}+J_{Y}^{b}J_{Z}^{c}J_{X}^{a}\\\nonumber &+J_{Z}^{c}J_{X}^{a}J_{Y}^{b}+J_{Z}^{c}J_{Y}^{b}J_{X}^{a}].
 \end{eqnarray}
 Therefore the mixed moments in this scheme are given by 
 \begin{equation}
  \mu^{abc}=\frac{1}{3!}Tr[\rho S_{MH} \{ J_{X}^{a}J_{Y}^{b}J_{Z}^{c}\}]
\end{equation}
where $S_{MH}\{\}$ stands for the  Margenau-Hill  symmetrizer as given on the right hand side of Eq.(32). Therefore, the Marganeau-Hill characteristic function can be written as,
\begin{eqnarray}
\Phi_{MH}=&\frac{1}{3!}Tr(\rho [ e^{iI_{x}J_{X}} e^{iI_{y}J_{Y}}e^{iI_{z}J_{Z}}+e^{iI_{x}J_{X}}e^{iI_{z}J_{Z}}e^{iI_{y}J_{Y}}+\\\nonumber 
&e^{iI_{y}J_{Y}}e^{iI_{x}J_{X}}e^{iI_{z}J_{Z}}+e^{iI_{y}J_{Y}}e^{iI_{z}J_{Z}}e^{iI_{x}J_{X}}+e^{iI_{z}J_{Z}}e^{iI_{x}J_{X}}e^{iI_{y}J_{Y}}\\ \nonumber
&+e^{iI_{z}J_{Z}}e^{iI_{y}J_{Y}}e^{iI_{x}J_{X}}])
\end{eqnarray} 
 
 Similarly, the Wigner-Weyl correspondence rule is given by Eq.(23) or equivalently by Eq.(24) and the characteristic function by Eq.(22). Thus the mixed moments in this scheme can be computed using Eq.(20).
 
 Note that the classical variates $XYZ$ have the same diagonal form as $J_{x},J_{y},J_{z}$, that is,
 
 ${\tiny
 \left( \begin{array}{cccccc}+j & & & & &\\ &(j-1)& &  & & \\& & \cdot& & &\\ & & & \cdot & &\\& & & & -(j-1)& \\ & & & & &-j\end{array} \right) \tiny}  $
 and the space of these random vairables has $(2j+1)^{3}$ points.
 Because of the algebra of spin matrices there are only $(2j+1)^{2}-1$ independent moments and all the others can be expressed in terms of these moments only.

 If $P(m_{x}m_{y}m_{z})$ is the probability mass function, then 
by definition the multivariate moments are given by,

 \begin{equation}
 \mu^{\alpha \beta\gamma}(XYZ)=\sum_{m_{x}}\sum_{m_{y}}\sum_{m_{z}}m_{x}^{\alpha}m_{y}^{\beta}m_{z}^{\gamma}P(m_{x}m_{y}m_{z}).
 \end{equation}
 Here, $\alpha, \beta,\gamma=0,1,2,....2j$ and $m_{x}, m_{y},m_{z}$ are the values taken by the random variables $X,Y,Z$ respectively. Since $ X,Y,Z$ are the classical random variables corresponding to the quantum mechanical operators $J_{X},J_{Y},J_{Z}$, for a spin-$j$ system, $m_{x}, m_{y},m_{z}=+j,.....,-j.$
Eq.(35) can be written as 
 \begin{equation}
  \mu^{\alpha \beta\gamma}(XYZ)=\sum_{m_{x}} m_{x}^{\alpha} {\cal A}_{m_{x}}^{\beta \gamma}
  \end{equation}
    and explicitly ,
  \begin{equation}{\tiny
  \left( \begin{array}{c}
  \mu^{0\beta\gamma}\\
  \mu^{1\beta\gamma}\\
  \cdot \\
  \cdot \\
  \mu^{2j\beta \gamma} \end{array} \right)
  =\left( \begin{array}{cccccc}
  1 & 1 &\cdot &\cdot & 1 & 1\\
  j& j-1 & \cdot & \cdot & -(j-1) & -j\\
  j^{2}& (j-1)^{2} & \cdot & \cdot & \cdot & (-j)^{2} \\
  \cdot & \cdot & \cdot & \cdot & \cdot & \cdot \\ 
  \cdot & \cdot & \cdot & \cdot & \cdot & \cdot \\
  j^{2j} &(j-1)^{2j} & \cdot & \cdot & \cdot &(-j)^{2j} \end{array}
  \right) 
  \left( \begin{array}{c}
  {\cal A}^{\beta\gamma}_{j}\\ 
  
  {\cal A}^{\beta\gamma}_{j-1}\\
  \cdot \\
  \cdot \\
  \cdot \\
  {\cal A}^{\beta\gamma}_{-j}\\ \end{array}
  \right)}
  \end{equation}
  \begin{equation}
  {\bf \mu}={\cal V}{\cal A}
  \end{equation}
 $\cal V$ is the well known  Vandermonde matrix.  This equation can be inverted to get ${\cal A}_{m_{x}}^{\beta \gamma}$ in terms of the known moments $\mu^{0\beta\gamma},\mu^{1\beta\gamma},\cdot,\cdot, \mu^{2j\beta\gamma}$ . But
 \begin{equation}
 {\cal A}_{m_{x}}^{\beta\gamma} = \sum_{m_{y}}m_{y}^{\beta} {\cal B}_{m_{x}m_{y}}^{\gamma}
 \end{equation}
 where
 \begin{equation}
 {\cal B}_{m_{x}m_{y}}^{\gamma}= \sum_{m_{z}}(m_{z})^{\gamma} P(m_{x}m_{y}m_{z}).
 \end{equation}
 Thus $P(m_{x}m_{y}m_{z})$ can be computed easily by successive inversions.
Observe that different sets of moments yield different probability distributions.
 
\section{Trivariate probability mass function for spin-$1$ system}

In the case of spin-$1$ matrices, it has been shown by Weaver \cite{weaver} that a particularly useful set of algebraic relations arises when one defines the set of matrices,$\Sigma_{1},\Sigma_{2},\Sigma_{3}$ to be
 \begin{eqnarray}
 \Sigma_{1}=&J_{x}^{2}-J_{y}^{2}\\ \nonumber
 \Sigma_{2}=&J_{x}J_{y}+J_{y}J_{x}\\ \nonumber
 \Sigma_{3}=&J_{z}.
 \end{eqnarray}
 Using the relation
 \begin{equation}
 J_{i}J_{j}J_{k}+ J_{k}J_{j}J_{i}=\delta _{ij}J_{k}+\delta _{jk}J_{i}
 \end{equation}
 the algebra of $\Sigma_{i}$'s has been shown to be 
 \begin{eqnarray}
 \Sigma_{1}^{2}=\Sigma_{2}^{2}=\Sigma_{3}^{2}=&J_{z}^{2}\\ \nonumber
 \Sigma_{i}\Sigma_{j}+ \Sigma_{j}\Sigma_{i}=&2\delta_{ij} J_{z}^{2}\\ \nonumber
 \Sigma_{i}\Sigma_{j}- \Sigma_{j}\Sigma_{i}=&2i\delta_{jk}\Sigma_{k}
 \end{eqnarray}
which combine to give
\begin{equation}
\Sigma_{i}\Sigma_{j}=\delta_{ij}J_{z}^{2}+i\epsilon_{ijk}\Sigma_{k} 
\end{equation}
exactly same as the Pauli spin matrix algebra. Here $\epsilon_{ijk}$ represents the Levi-Civita symbol.

Since X,Y,Z are given by the diagonal matrix ${\tiny
 \left( \begin{array}{ccc}+1 & &\\ &0&  \\ & & -1\end{array} \right) \tiny}$, the space of our random variables have only 27 points and we need exactly 27 moments $\mu^{ijk};i,j,k=0,1,2$ to compute the probability mass function.  

Using the relations (41) to (44) extensively together with 
\begin{equation}
Tr[\rho(J_{x}^{2}+J_{y}^{2}+J_{z}^{2})] =2,
\end{equation} it can be shown that, of the 27 moments, only 8 are independent.
We choose the independent moments to be $\mu^{001},\mu^{010},\mu^{100},\mu^{200}\mu^{020},\mu^{101},\mu^{110},\mu^{011}$. The remaining 19 moments can be computed interms of the above 8 independent moments.

Thus the 27 points of the probability mass function both in Margenau-Hill and Wigner-Weyl prescriptions can be obtained  interms of the 8 independent moments as
\begin{equation}
P_{MH}(m_{x}m_{y}m_{z})=\frac{1}{48}\sum_{i=x,y,z} m_{i} Tr[\rho {\bf J_{i}}] +\frac{1}{48} \sum_{k\neq i=x,y,z} m_{i}m_{k}Tr[\rho({\bf J_{i}}{\bf J_{k}}+{\bf J_{k}}{\bf J_{i}})],
\end{equation}
when all the 3 arguments are non-zero
\begin{eqnarray}
P_{MH}(m_{x}m_{y}m_{z})=&\frac{1}{12}\sum_{i=x,y,z} m_{i} Tr[\rho {\bf J_{i}}]+\frac{1}{4}[\sum_{i=x,y,z} m_{i}^{2} Tr[\rho {\bf J_{i}^{2}}] -1]\\ \nonumber & + \frac{1}{48}\sum_{k\neq i=x,y,z} m_{i}m_{k} Tr [\rho ({\bf J_{i}}{\bf J_{k}}+{\bf J_{k}}{\bf J_{i}})],
\end{eqnarray}
when two of the arguments are non-zero,
\begin{equation}
P_{MH}(m_{x}m_{y}m_{z}) =\frac{1}{12}\sum_{i=x,y,z} m_{i} Tr[\rho {\bf J_{i}}]
\end{equation}
when two of the arguments are zero and
\begin{equation}
P_{MH}(m_{x}m_{y}m_{z})=0
\end{equation}
where $m_{x}=m_{y}=m_{z}=0.$ 

Wigner-Weyl probability mass function turns out to be
 \begin{equation}
P_{WW}(m_{x}m_{y}m_{z})=\frac{1}{120}\sum_{i=x,y,z} m_{i}Tr [\rho {\bf J_{i}}] +
\frac{1}{96}\sum_{k\neq i=x,y,z} m_{i}m_{k} Tr[\rho ({\bf J_{i}}{\bf J_{k}}+{\bf J_{k}}{\bf J_{i}})]
\end{equation}
when all the 3 arguments are non-zero,
\begin{equation}
P_{WW}(m_{x}m_{y}m_{z})=\frac{1}{15}\sum_{i=x,y,z} m_{i}Tr [\rho {\bf J_{i}}] +
\frac{1}{12}\sum_{k\neq i=x,y,z} m_{i}m_{k} Tr[\rho ({\bf J_{i}}{\bf J_{k}}+{\bf J_{k}}{\bf J_{i}})]
\end{equation}
when two of the 3 arguments are non-zero,
\begin{equation}
P_{WW}(m_{x}m_{y}m_{z})=\frac{1}{5}\sum_{i=x,y,z} m_{i}Tr [\rho {\bf J_{i}}] +
\frac{5}{12}\sum_{k\neq i=x,y,z} m_{i}^{2} Tr[\rho ({\bf J_{i}}^{2})]-\frac{1}{6}
\end{equation}
when two of the 3 arguments are zero and
\begin{equation}
P_{WW}(m_{x}m_{y}m_{z})=\frac{-1}{3}
\end{equation}
where $m_{x}=m_{y}=m_{z}=0.$ 

Notice that 
\begin{equation}
\sum_{m_{x},m_{y},m_{z}} P_{MH}(m_{x}m_{y}m_{z})=1.
\end{equation}
and
\begin{equation}
\sum_{m_{x},m_{y},m_{z}} P_{WW}(m_{x}m_{y}m_{z})=1.
\end{equation}

 From the expressions for the probability mass function it is evident that they can assume negative values also.
Summing over any one of the indices in the expressions for probability distributions one can obtain the biavriate probability mass function which is capable of describing the aligned, biaxial spin-1 system which can be generated by exposing a spin-1 nucei with non-zero electric quadrupole moment to external electric quadrupole field generated by a suitable crystal lattice \cite{gr4}.
Summing over any of the two indices we can get the univariate probability distribution.  As expected both Margenau-Hill and Wigner-Weyl distributions lead to the same univariate distribution.  Such distributions of any spin-j system represent oriented spin systems\cite{blin}.  For instance an oriented assembly reults when a spin-j system is exposed to external dipole magnetic field.  Such probability distributions are semi positive definite and are nearest to the classical probability distributions.  Where as , the bivariate and trivariate distributions are generally quasi probabilistic and the corresponding spin assemblies are called non-oriented\cite{gr5}.  Thus any spin-j system has to belong to one of the mutually exclusive classes viz., oriented or non-oriented systems.  It is interesting to study the squeezing property of the univariate, bivariate and trivariate  spin-j systems.
\section{Spin squeezing}
One may consider the spin system to be squeezed in $J_{x}$ or $J_{y}$ provided
\begin{equation}
\Delta J_{x}^{2}\langle  \frac{1}{2}\vert Tr \rho J_{z} \vert
\end{equation}
or
\begin{equation}
\Delta J_{y}^{2}\langle  \frac{1}{2}\vert Tr \rho J_{z} \vert
\end{equation}
where $\Delta J_{y}^{2}$ and $\Delta J_{y}^{2}$ represent the variances in $J_{x}$ and $J_{y}$ respectively. It has already been noted by Kitagawa and Ueda\cite{kit} that this definition implies that a coherent spin state is squeezed if placed in an appropriate coordinate system. To overcome this difficulty, they consider a spin-j system to be made up of 2j elementary spinors and choose a coordinate system whose z-direction coincides with the mean spin direction.

For a mixed system, the vector polarization is defined as
\begin{equation}
\vec P =  \frac{Tr \rho \vec J}{Tr \rho}.
\end{equation}
Kitagawa and Ueda\cite{kit}  \quad refer  to this as the mean spin direction.\\
The spherical components of vector polarization $\vec P$ are related to $t_{q}^{1}$ through
\begin{equation}
P^{1}_{q}= \sqrt{\frac{j(j+1)}{3}} t^{1}_{q}.
\end{equation}
Choosing the Z-axis along $\vec P$ one is left with infinite choices of $X$ and $Y$ axes.  Now make a rotation about $\vec P$ such that the covariance 
\begin{equation}
(\Delta J_{x}J_{y})^{2}= \frac{Tr (\rho (J_{x}J_{y}+J_{y}J_{x}))}{2}-Tr (\rho J_{x}) Tr( \rho J_{y})=0.
\end{equation}
In this resulting frame,  the Schrodinger inequality\cite{hrob} 
\begin{equation}
(\Delta J_{x})^{2} (\Delta J_{y})^{2} (-\Delta J_{x}J_{y})^{2}\geq  \frac{\vert Tr (\rho ([J_{x},J_{y}]))\vert^{2}}{4}
\end{equation}
and the Heisenberg-Robertson inequality\cite{hrob},
\begin{equation}
(\Delta J_{x})^{2} (\Delta J_{y})^{2} \geq \frac{\vert Tr(\rho J_{z})\vert ^{2}}{4}
\end{equation}
are of the same form and the definition of spin squeezing is unambiguous ie.,
\begin{equation}
(\Delta J_{x})^{2} <  \frac{\vert Tr (\rho J_{z})\vert}{2}
\end{equation}
or
\begin{equation}
(\Delta J_{y})^{2} <  \frac{\vert Tr (\rho J_{z})\vert}{2}.
\end{equation}
In terms of spherical tensor notation, the squeezing criterion becomes
\begin{equation}
1+\left [\frac{3(2j+3)(2j-1)}{j(j+1)40}\right ]^{1/2}(2t_{2}^{2}-\sqrt{\frac{2}{3}}t_{o}^{2}) < \frac{1}{2}\left [\frac{3}{j(j+1)}\right ]^{1/2} \vert t_{o}^{1}\vert.
\end{equation}
This condition is never satisfied by oriented systems\cite{ksm}, in other words univariate systems.
One has to consider bivariate or trivariate systems.
The simplest bivariate system\cite{ss} one could think of is a spin-j assembly of nuclei with non-zero electric quadrupole moment exposed to external electric and quadrupole field generated by electrons in suitable crystal lattice site.  Here, $t_{0}^{1}=0$ and the squeezing criterion is never satisfied.
Therefore one needs to consider a trivariate system.
If the above bivariate system is exposed to an external magnetic field along the z-axis of the Principle Axes Frame(PAF) , the resultant spin assembly will need all the three variates $J_{x},J_{y},J_{z}$ for its description\cite{ss}.
In the  case of spin-1 assembly , there are situations when the assembly exhibits squeezing and a few values are presented in table.1.

\begin{center}
{\bf Table 1: Trivariate,squeezed spin-1 system specified by
its non-zero spherical tensor parameters}
\vskip0.3in
\begin{tabular}{ccccc}
\hline 
$t_{o}^{1}$& $t_{o}^{2}$ & $t_{2}^{2}$&$(\Delta J_{x})^{2}$&${\vert Tr (\rho J_{z} ) \vert \over 2}$\\
\hline 
 -0.7506 & 0.495 & -0.4453 & 0.2929 & 0.3064 \\
 \hline
 -0.6298 & 0.4737 & -0.5307 & 0.2486 & 0.2571\\
 \hline
 -0.7506 & 0.4526 & -0.4453 & 0.3028 & 0.3064 \\
 \hline
 \end{tabular}
 \end{center}
\section{CONCLUSIONS}
In this paper it has been shown that any general spin-$j$ 
assembly can be characterised by a trivariate probability 
distribution function with $J_{x},J_{y},J_{z}$ as the variates.
The distribution needs $n^{2}-1$ independent moments of  
$J_{x},J_{y},J_{z}$ for its complete description. Because of 
non-commutativity of these variates there exist various 
correspondence rules which associate different quantum mechanical
expressions to the same classical expression for moments.
Using two such correspondence rules viz, Wigner-Weyl and Margenau-Hill
rules, we have computed the probability mass function to arbitrary spin-$1$ 
systems. These distributions are in general found to be quasi-probabilistic.
Starting from moments we have also outlined the method to obtain 
the probability mass function by the method of inversion. Since univariate 
distributions correspond to oriented spin assemblies, they are never squeezed.
Thus one has to look for signatures squeezing in the bivariate or trivariate distributions.
In the case of spin-$1$ assembly, the trivariate distribution which describes a spin-$1$
nucleus like $H^{2},N^{14}$ exposed to combined electric and magnetic
field shows squeezing in certain cases.
\section*{Acknowledgements}
The author thanks University Grants Commission(India) and Indian
National Science Academy for providing partial financial
assistance for attending ICSSUR9 at Besancon, France and presenting this paper.

\end{document}